\documentclass[sigconf,10pt]{acmart}

\renewcommand\footnotetextcopyrightpermission[1]{}
\usepackage{xspace}
\usepackage{graphicx}
\usepackage{marvosym}
\usepackage{ifthen}
\usepackage{pifont}
\usepackage{paralist}
\usepackage{enumitem}
\usepackage[abbreviations]{foreign}  
\usepackage{hyperref}
\usepackage{caption}
\usepackage{subcaption}

\newboolean{showcomments}
\setboolean{showcomments}{false}
\ifthenelse{\boolean{showcomments}}
{\newcommand{\mynote}[3]{
    \fbox{\bfseries\sffamily\scriptsize#1}
    {\small$\blacktriangleright$\textsf{\emph{\color{#3}{#2}}}$\blacktriangleleft$}}}
{\newcommand{\mynote}[3]{}}

\newcommand{\clflush}{\texttt{clflush}\xspace}
\newcommand{\clwb}{\texttt{clwb}\xspace}
\newcommand{\clflushopt}{\texttt{clflushopt}\xspace}
\newcommand{\sfence}{\texttt{sfence}\xspace}
\newcommand{\mfence}{\texttt{mfence}\xspace}
    
\newcommand{\mm}[1]{\mynote{mm}{#1}{blue}}

\newcommand{\patcher}{\textsc{Bentō}\xspace}

\begin{document}

\begin{abstract}
Persistent Memory (PM) is a new storage technology that brings high performance, byte addressability, and persistency for a lesser cost than DRAM.
Due to cache volatility and store reordering, developers must use explicit instructions (e.g.: flush and fence) to guarantee that the application state remains consistent upon crashes.
This is difficult to get right and, in fact, several tools have been created to detect bugs in PM programs.
To overcome this difficulty, programmers tend to be overly conservative, for instance, by enforcing unnecessary ordering constraints, which partially forfeits the performance benefits of using PM. 

In this paper, we study the impact that different combinations of persistency instructions have in several PM programs and found that a specific combination can lead to performance improvements while preserving the original crash-consistency semantics.
Based on these results we developed \patcher an automatic and black-box binary rewriter that can boost the performance of existing PM programs by up to 15\% with minimal programmer effort.

\end{abstract}

\pagestyle{plain}

\title{\patcher: Optimizing Persistent Memory Programs}
\author{Sebastião Amaro, João Gonçalves, Miguel Matos}
\affiliation{%
  \institution{IST, Universidade de Lisboa, INESC-ID}
}
\email{{sebastiao.amaro,joao.tiago.goncalves,miguel.marques.matos}@tecnico.ulisboa.pt}

\renewcommand{\shortauthors}{S. Amaro et al.}

\maketitle

\sloppy

\section{Introduction \& Related Work}

Persistent Memory (PM) is a new storage technology that is byte-addressable and aims to bring performance close to DRAM for a lesser cost (per GB) while providing data persistency upon program crashes and restarts. 
PM has been used in a wide range of programs such as PM-specific indexes, PM filesystems and databases such as Microsoft SQL Server~\cite{sql}, memcached~\cite{memcached} or Redis~\cite{redis}.

The combination of byte-addressability and data persistency opens new opportunities for developers to design their programs.
However, since modern CPU architectures buffer and reorder store operations, and CPU caches are volatile and have unpredictable cache eviction policies, there are no guarantees that the state is persisted in a crash-consistent fashion\footnote{
eADR~\cite{eadr} can alleviate this problem by extending the persistency domain to CPU caches but: i) stores can still be reordered (see \S\ref{sec:background}) and ii) its additional costs make it unlikely to become widely available and hence programs must still work properly in the absence of eADR.}.
Therefore developers must explicitly use flush instructions and memory fences to ensure that the state is persisted in such an order that, should a crash occur, the state remains consistent upon a restart.
Moreover, there are several variants of flush and fence instructions, which offer different performance and semantics trade-offs.

Overall, selecting and using the right instructions that maximize performance and guarantee correctness is a complex and error-prone task as shown by the growing body of PM bug detection tools~\cite{pmtest,pmdebugger,yat,jaaru,agamotto,xfdetector,mumak,witcher,durinn,pmrace,hawkset}.
This complexity can lead developers to be more conservative than necessary, such as employing more flushes and fences and with stronger semantics than needed, or imposing unnecessary ordering restrictions, which can forfeit the performance benefits of using PM in the first place.

There have been multiple studies
\cite{perfomancemeasurespm,idiosyncrasiesPM,pmdimmbuffering,unexpectedperfomanceofinteloptane,earlystudyintel,lessonslearnedpmdbms,evaluationofinteloptane,evalindexes,pmioprimitives,gordon,modelingnvm,pactree} with the goal of profiling the performance of PM, understanding its design, and how it affects performance. 
\citeauthor{pmdimmbuffering}~\cite{pmdimmbuffering} performed an in-depth study of the read and write buffering, read and write amplifications, the overhead of different types of persistence barriers, and the pros/cons of different persistency models in Intel Optane DC. 
PerMA-BEnch~\cite{permabench} is a configurable benchmark framework to evaluate the bandwidth, latency, and operations per second for customizable database-related PMem access. 
This extensive evaluation provided eight general and implementation-specific insights that influence PM performance.
\mm{se sobrar espaço podemos adicionar mais alguns dos sistemas comentados}

In short, the existing studies focused mostly on the read/write performance of the PM devices themselves, and not on the impact of the different PM instructions on program performance and correctness in concrete and representative workloads.
For that reason, and to better understand this issue, we performed a study with real applications and representative workloads, comparing the performance of different combinations of instructions on program performance and consistency (\S\ref{sec:study}).
We found that a specific combination, which is generic and can be applied without knowledge of the program semantics, can lead to improvements up to 15\% and, in the worst case, maintain the original performance. 

Motivated by these results, we developed \patcher, a tool that automatically and in a black-box fashion rewrites the program binary with the instructions that lead to the best performance without changing the existing crash-consistency guarantees (\S\ref{sec:patching}).

Finally, we launch a discussion on possible future directions, namely on finding the \textit{Minimal Crash-Consistency State} of a program, \emph{i.e.} the point at which the program is crash consistent with the minimal amount of flush and fence instructions, which corresponds to the best possible trade-off between correctness and performance (\S\ref{sec:discussion}).


\section{Background}
\label{sec:background}
In this section, we provide background on the persistence semantics of PM.
For simplicity, we focus on Intel x86
~\cite{pmsematincs} but other architectures such as ARM also have similar persistency semantics and offer instructions with varying guarantees.
Intel x86 follows a relaxed, buffered persistency model. 
The relaxed property causes write operations to not be persisted in the same order they become visible to other threads.
 The buffered property relates to when persistent writes occur, in a buffered architecture they can be asynchronous and writes can be queued while the execution proceeds. 
When a store occurs, it is first queued in a store buffer and eventually reaches the CPU cache.
It will only be persisted when it reaches the Write Pending Queue (WPM) in the PM controller.
The store can reach the WPQ in an unpredictable way since the cache evicts lines when it needs to load new data, and simply relying on this behaviour will, naturally, result in inconsistent states in case of a crash.
Therefore, developers must use special instructions to control in which order the stores reach PM. These instructions 
can be divided into two groups: flushes and fences.
Regarding flushing instructions, Intel x86 offers \clflush, \clflushopt and \clwb. \clflush evicts and persists a single cache line, and cannot be reordered with other store and flush instructions.
\clflushopt evicts and persists a single cache line but can be reordered until a fence instruction appears.
\clwb persists a single cache line, however, it does not evict it from the cache and can be reordered as well.
Two fencing instructions exist.
\mfence guarantees that all buffered loads, stores, and flushes before the fence are executed before the ones after the fence, while \sfence only provides this guarantee for stores and flushes. 
In a sense, we can establish a parallel between using these instructions and the use of concurrency primitives.
On the one hand, not using them at all, or placing them in the wrong places, will result in incorrect programs.
On the other hand, using these instructions more than strictly necessary, or using unoptimized counterparts, will penalize performance.

\section{Performance Study}
\label{sec:study}

In this section, we study the cost of the different combinations of flush and fence instructions and discuss their impact on the crash-consistency guarantees.
All experiments were performed on a 128-core Intel(R) Xeon(R) Gold 6338N CPU @ 2.20GHz, with 256 GB of RAM, and 1 TB Intel DCPMM in App Direct mode running Ubuntu 22.04 with kernel 5.15.
Moreover, for all experimental results, we present the average and standard deviation of five independent executions.

\begin{figure}
	\includegraphics[width=1.0\linewidth]{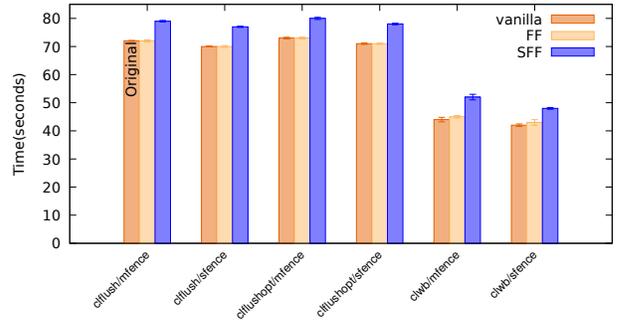}
	\caption{Performance impact of different combinations of flush and fence instructions in a Level Hashing workload with 100\% inserts.}
	\label{plot:lhscenarios}
\end{figure}

\subsection{Preliminary Study}\label{preliminary}
We start by studying the performance impact of each flush/fence combination in Level Hashing~\cite{level_hashing}.
Level Hashing is a write-optimized and high-performance hashing index for PM.
In terms of implementation, Level Hashing uses three \clflush instructions, two of which are followed by an \mfence.
Our goal is to answer the following two questions: 
i) what is the performance impact of following a simple, conservative, approach to crash consistency? and
ii) which combination of flush and fence instructions provides the best performance for similar crash-consistency guarantees?
We ran a workload with 100 million inserts (100\% writes) and measured the total time taken to complete the workload.
The results are depicted in Figure~\ref{plot:lhscenarios}.
We consider three variants of LevelHashing:
i) \textit{vanilla} which corresponds to the original implementation,
ii) \textit{FF} a variant where each flush is followed by a fence, and
iii) \textit{SFF} a variant where each store is followed by a flush and a fence.
The \textit{vanilla} variant represents a scenario where the developer is judicious about the use of PM instructions, while the \textit{FF} and \textit{SFF} represent  two progressively conservative scenarios where the developer simplifies the problem by following simple ideas.
We then manually adjusted each variant by implementing the different combinations of flush and fence instructions, resulting in a total of 6 combinations (3 flush instructions times 2 fence instructions).

To answer the first question, we can conclude that the performance of adding either an \mfence or an \sfence after every flush (\textit{FF} variant) does not significantly hinder performance.
This is because the original Level Hashing already had two of the three \clflush operations followed by a fence, which means we only added an extra fence.
However, doing a flush after every store (\textit{SFF} variant) significantly hinders the performance, which is to be expected since it eliminates the potential for batching the stores.

To answer our second question, 
we compare the results for the different combinations of persistency instructions.
As we can observe, using \texttt{clwb} with any of the fence instructions in any of the variants had much better results than their counterparts.
The major difference between \texttt{clwb}, \texttt{clflush} and \texttt{clflushopt} is that the former retains the cache line it wrote to memory. 
This could aid future operations because we might spare the application from accessing PM if a value is already in the cache.
At first sight, and given that the workload is 100\% writes, it might seem that invalidating the cache lines by using \clflush or \clflushopt would have little to no effect when compared to preserving the cache lines when using \clwb.
However, LevelHashing metadata management in operations such as resizing implies reading existing values which benefits from cache locality.
Given this insight, combined with the knowledge that reading from PM is a major overhead when compared with reading from the cache, we decided to further study the proposition that \texttt{clwb} should be used over \texttt{clflush} and \texttt{clflushopt} to achieve better performance in PM applications. 

\subsection{Flush Performance Impact}
\label{sec:flushstudy}

In this section, we applied the insights obtained in our preliminary study, extending them to other PM datastore applications.
Besides Level Hashing, we analyzed FAST\&FAIR~\cite{FASTFAIR} and WOART~\cite{woart}, which are PM-optimized B+-tree and radix-tree implementations, respectively.
All three applications use \clflush for persisting stores but, as we noted in \S\ref{preliminary}, \clwb has considerably better performance.
However, its weaker semantics make it so that a direct swap does not preserve the correctness guarantees.
With that in mind, we studied the impact of replacing the original \clflush with a \clwb followed by an \sfence which effectively preserves the crash-consistency guarantees of the original program.
We performed this change manually for each application and, in the cases where the original code already has a \clflush followed by a fence, we simply replace the \clflush with a \clwb to avoid having redundant fence operations.

\begin{figure*}
    \centering
    \includegraphics[trim={1cm 0.4cm 0.5cm 1cm},clip,width=\textwidth]{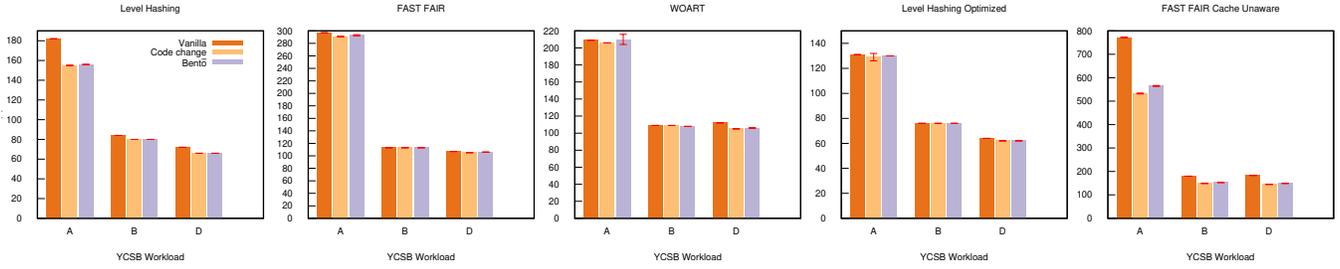}
	\caption{Performance impact of different flush and fence combinations for YCSB workloads A, B and D. The vanilla bars stands for the original unmodified application (which use \clflush), manual is our manual modifications (using \clwb followed by an \sfence) described in \S\ref{sec:flushstudy}, and \patcher is the automatic rewriting proposed in \S\ref{sec:patching}.}
     \label{plot:ycsb}
\end{figure*}

The workload of the previous section had only insert operations and served only as a preliminary study.
In this section, we use more realistic workloads by relying on the well-known YCSB benchmark~\cite{ycsb}.
We use the following workloads: 
workload A which consists of 50\% reads and 50\% updates; 
workload B which consists of 95\% reads and 5\%  updates; 
and workload D which consists of 95\% reads and 5\% inserts but where reads access the most recently inserted keys. 
All workloads follow a \textit{zipfian} distribution, perform 25 million operations, and are preceded by a load phase that pre-fills the datastore with 25 million keys.
We ignored workload C since it is 100\% reads and hence performs no flushes or fences, and workload E as ranges are not supported by the datastores.
For workload F we obtained results similar to workload A (not shown).
The results are presented in Figure~\ref{plot:ycsb} where the bars labeled vanilla represent the original unmodified applications and the manual bars represent our modifications discussed above.
Next, we discuss the results for each application.


\subsubsection{Level Hashing}
\label{levelhashing}
For workload A, substituting \clflush with \clwb plus \sfence led to a performance improvement of 15\%.
Each update and insert operation in Level Hashing will have three cache line flushes, which are performed in a sequence after the respective stores are performed.
It flushes the key, value, and token (which indicates whether a bucket is empty or not) individually.
The three values are implemented using \texttt{uint64\_t} (8 bytes) and are consecutive in memory, meaning they fit in a single cache line (64 bytes total size in 64-bit systems).
\clflush invalidates the cache line it acts upon, leading to an eviction.
This means that after flushing the key, the cache line (where the value and token are also collocated) is evicted. Subsequently, to flush the value it must be first loaded back from PM incurring, in additional overhead, and the same applies for the token.
By changing \clflush with \clwb, the cache line is no longer evicted, meaning we do not incur the overhead of having two extra reads from PM. We believe this to be the major reason for the improvement we had.
Given that a flush acts on the whole cache line, and considering that the triplet (key,value,token) fits in a single cache line, in fact, a single flush would suffice to flush the whole triplet.
To confirm this, we instrumented the application to trace all the performed flushes and indeed confirmed that two redundant flushes were performed for each update.
Since this impacts performance, we decided to also consider an optimized version, \textit{Level Hashing Optimized}, in which we manually fixed the issue.
As anticipated, the potential improvements by swapping the \clflush for a \clwb followed by a \sfence disappear since the behavior previously described no longer occurs, and there is no correlation between reads and write, thus we do not take advantage of leaving values in the cache.

For workload B, we observed a 5\% performance improvement.
This is due to the fact that reads do not result in writes to PM and consequently do not result in flushes either. 
We still bring a performance improvement but to a lesser extent given the nature of the workload.
The Level Hashing Optimized version, for the same reasons as workload A, does not benefit from the manual modifications.

Finally, for workload D we observed an 8\% improvement.
This workload consists of mostly reads (95\%) and hence the datastore does not frequently show the behavior described above.
However, since the workloads exhibit a correlation between reads and inserts, where reads access recently inserted keys, the impact of cache line invalidation is still relevant.
By using \clwb to persist the value of a key, the cache holds the value and is thus able to respond to read operations on the same key without having to fetch the value from PM.
This applies to both the regular Level Hashing and our optimized version, although the improvements in the optimized version were smaller for the reasons discussed above.



\subsubsection{FAST\&FAIR}

\label{fastfair}
Failure-Atomic ShifT, and Failure-Atomic In-place Rebalance, is a PM-optimized B+-tree implementation.
Note that FAST\&FAIR does not have an update operation since it allows for duplicate keys.
Therefore we treated updated operations as inserts.
Regarding workload A, we observed a modest performance increase.
In detail, B+-tree inserts values by shifting their neighbors.
On average, each time a new value is inserted half of its neighbors are shifted.
Since values are an entry of 8 bytes and a pointer of 8 bytes, thus 16 bytes, 4 values fit in the same cache line.
With that in mind, when shifting, FAST\&FAIR only performs a \clflush when it crosses a boundary to the next cache line.
We note that this optimization is directly tied to FAST\&FAIR's design and is not trivially deduced nor applicable.
To mimic a more conservative developer and/or less aware of the PM cache behavior, we developed a hand-tailored conservative version that flushes after every store in the key insertion phase which we call \textit{FAST\&FAIR Cache Unaware}.
Similarly to Level Hashing, this results in redundant flushes.
As discussed earlier, since \clflush evicts the cache line, this also results in three extra data fetches from PM.
When using \clwb these extra PM reads will not happen, causing a 30\% improvement in performance.



Workload B shows negligible performance variance in the original FAST\&FAIR, and we attribute the same reasons as with Level Hashing. 
For FAST\&FAIR Cache Unaware, we observe a 17\% performance improvement because even though the portion of writes is small (5\%) these inserts have a high cost. 

Finally, for workload D we observe a small improvement in FAST\&FAIR.
This is because FAST\&FAIR has few \clflush instructions and therefore there is a smaller room for improvement.
In the case of FAST\&FAIR Cache Unaware, because the number of flushes is three times larger we observe a performance improvement of 20\%.
This is also explained by the locality of reads and inserts which further benefits from fewer cache evictions.


\subsubsection{WOART}
Write Optimized Adaptive Radix Tree, is a PM-optimized radix tree.
As with FAST\&FAIR, the update operation does not exist, thus we use the insert operation instead.
One of the key properties of a radix tree is that to find a specific key it does not need to perform key comparisons.
This causes less cache pollution, and hence fewer cache misses, than its B-tree counterparts. 
Moreover, and unlike FAST\&FAIR, WOART is not cache line aware meaning it might perform redundant flushes.
Regarding the results, for workload A we did not see substantial improvements because reads are not related to inserts and hence not evicting the cache does not bring any advantage.

For workload B we also did not observe significant changes for the same reasons of Level Hashing and FAST\&FAIR.

In workload D, however, we observe a 6\% improvement which we attribute to the fact that we read recently inserted values and hence leverage the cache.
One might ask why WOART performs this way and FAST\&FAIR does not.
We point out two reasons. 
Firstly, since FAST\&FAIR is cache-aware it performs fewer cache line flushes and hence leaves less room for the improvement brought by replacing a \clflush instruction with a \clwb followed by an \sfence. 
Secondly, since WOART has less cache pollution,  \clwb leaves values in the cache for longer periods.
This is extremely relevant because in this workload we know that a read operation in the near future will be performed in the recently inserted value.


\section{\patcher}
\label{sec:patching}

The findings of the previous section show that it is possible to improve the performance of applications by up to 15\% in certain workloads without incurring performance penalties in adversarial workloads.
This is achieved by replacing \clflush instructions with \clwb followed by an \sfence.
By not evicting the cache when flushing, this allows applications to take advantage of cache locality while preserving the original crash-consistency guarantees.
This motivated us to build a tool, \patcher, to perform this substitution automatically and in a black-box manner.
Our goal is to allow developers to use \patcher without requiring a deep understanding of the low-level details of how each PM instruction works, and without requiring changes to the source code.
This enables developers to program more defensively and, hopefully, construct PM programs with stronger crash-consistency guarantees that can still perform well and take advantage of the underlying hardware characteristics.


We analyzed several alternatives to achieve this and ended up opting for binary rewriting, in contrast with, for instance, a compiler pass, since it is black-box (no dependence on source code) and ergonomic, in the sense that it does not impact the normal build pipeline of target applications.
\patcher relies on e9patch~\cite{e9patch}, a tool that can statically rewrite x86\_64 binaries without any knowledge of the control flow information.
It leverages a combination of instruction punning, padding, and eviction to insert jumps to trampolines  without the need to move other instructions and perform the expensive jump target re-computation in $99\%$ of the cases.
Briefly, \patcher receives as input a binary and outputs a patched binary with all \clflush instructions replaced by a \clwb followed by an \sfence.
Because the original binary might already have fence instructions, blindly adding an \sfence after each \clwb might result in unnecessary overhead.
With that in mind, \patcher's patching rules additionally check if the subsequent instruction to the \clflush is either a \sfence or \mfence.
If that is the case we only replace the \clflush with a \clwb and do not add the additional \sfence to prevent additional overhead.
It is important to note that this does not cover all cases, since the flush and fence instructions might be separated by an arbitrarily large number of other instructions which cannot be captured by the matching rules of our binary rewriting tool.
This could be addressed by performing static binary analysis which we leave for future work.


The results for \patcher are depicted in Figure~\ref{plot:ycsb}.
As it is possible to observe, and despite the limitation discussed above, \patcher achieves almost identical results to the manual approach we discussed in the previous section.
We, therefore, believe that developers can use \patcher with very little effort to improve the performance of their PM applications while preserving the original crash-consistency guarantees.
\section{Discussion}
\label{sec:discussion}
PM combines near DRAM access latencies with disk-like durability for a lesser cost. 
However, to fully exploit the hardware characteristics developers must judiciously use fence and flush instruction.
Not using theses instructions in the right places, or not using them at all, will result in incorrect programs, whereas being excessively conservative and using them more than strictly necessary, or using their unoptimized counterparts, will hinder performance.

Our study showed that by selecting the right combination of flushes and fences, and more precisely a \clwb followed by an \sfence, it is possible to improve performance by up to 15\% in certain applications and workloads.
Interestingly, this is achieved without ever degrading performance, even in adversarial workloads that do not leverage the cache, and while preserving the original crash-consistency guarantees.
One thing to note is that we assume the original application is correct, i.e. that it ensures crash-consistency.
If that is not the case, this means that the application is missing further flush and fence instructions which, in turn, opens the door for further performance improvements.



Encouraged by the experimental results of our study, we developed \patcher, a tool that automatically rewrites the application binaries and achieves performance improvements almost identical to the manual source code rewriting we performed.
\patcher is automatic and black-box - it works by rewriting the application binary and does not require any access to the source code.
This allows developers to use \patcher to improve the performance of their PM applications with little effort and, to the best of our knowledge, no downsides.

We now discuss future research directions enabled by this work.
\patcher brings good performance improvements by exploiting a simple idea.
Naturally, more conservative approaches that flush and fence more than necessary can benefit more from the use of \patcher, while already optimized approaches leave less room for improvement.
Currently, we are exploring mechanisms to further improve the performance of PM applications without sacrificing correctness.
To achieve this we plan to study and develop the concept of Minimal Crash-Consistency State (MCCS) defined as the minimum set of persistency constraints needed that still guarantee crash-consistency.
To determine the MCCS, we need an oracle capable of determining whether the target application is crash-consistent or not.
As a first step, we could rely on existing PM bug detection tools~\cite{pmtest,yat,jaaru,agamotto,xfdetector} as an imperfect instance of this oracle.
Given such an oracle, and based on the knowledge of the performance and semantic trade-offs offered by the existing fence and flush instructions, we could follow a smart-iterative process that performs successive transformations over the binary until the MCCS is reached.

\section*{Acknowledgments}
This work was supported by Fundação para a Ciência e a Tecnologia (FCT) under plurianual grant UIDB/50021/2020,  PhD scholarship 2021.07401.BD, and research project grant PTDC/CCI-COM/4485/2021 (Ainur).

\bibliographystyle{ACM-Reference-Format}
\bibliography{bibliography}

@article{idiosyncrasiesPM,
author = {Gugnani, Shashank and Kashyap, Arjun and Lu, Xiaoyi},
title = {Understanding the Idiosyncrasies of Real Persistent Memory},
year = {2020},
issue_date = {December 2020},
publisher = {VLDB Endowment},
volume = {14},
number = {4},
issn = {2150-8097},
url = {https://doi.org/10.14778/3436905.3436921},
doi = {10.14778/3436905.3436921},
journal = {Proc. VLDB Endow.},
month = {dec},
pages = {626–639},
numpages = {14}
}

@article{perfomancemeasurespm,
  author    = {Joseph Izraelevitz and
               Jian Yang and
               Lu Zhang and
               Juno Kim and
               Xiao Liu and
               Amir Saman Memaripour and
               Yun Joon Soh and
               Zixuan Wang and
               Yi Xu and
               Subramanya R. Dulloor and
               Jishen Zhao and
               Steven Swanson},
  title     = {Basic Performance Measurements of the Intel Optane {DC} Persistent
               Memory Module},
  journal   = {CoRR},
  volume    = {abs/1903.05714},
  pages     = {18,25},
  year      = {2019},
  url       = {http://arxiv.org/abs/1903.05714},
  eprinttype = {arXiv},
  eprint    = {1903.05714},
  bibsource = {dblp computer science bibliography, https://dblp.org}
}

@inproceedings{pmdimmbuffering,
author = {Xiang, Lingfeng and Zhao, Xingsheng and Rao, Jia and Jiang, Song and Jiang, Hong},
title = {Characterizing the Performance of Intel Optane Persistent Memory: A Close Look at Its on-DIMM Buffering},
year = {2022},
isbn = {9781450391627},
publisher = {Association for Computing Machinery},
address = {New York, NY, USA},
url = {https://doi.org/10.1145/3492321.3519556},
doi = {10.1145/3492321.3519556},
booktitle = {Proceedings of the Seventeenth European Conference on Computer Systems},
pages = {488–505},
numpages = {18},
location = {Rennes, France},
series = {EuroSys '22}
}

@ARTICLE{unexpectedperfomanceofinteloptane,
  author={Mason, Tony and Doudali, Thaleia Dimitra and Seltzer, Margo and Gavrilovska, Ada},
  journal={IEEE Computer Architecture Letters}, 
  title={Unexpected Performance of Intel® Optane™ DC Persistent Memory}, 
  year={2020},
  volume={19},
  number={1},
  pages={55-58},
  doi={10.1109/LCA.2020.2987303}}

@inproceedings{earlystudyintel,
author = {Weiland, Mich\`{e}le and Brunst, Holger and Quintino, Tiago and Johnson, Nick and Iffrig, Olivier and Smart, Simon and Herold, Christian and Bonanni, Antonino and Jackson, Adrian and Parsons, Mark},
title = {An Early Evaluation of Intel's Optane DC Persistent Memory Module and Its Impact on High-Performance Scientific Applications},
year = {2019},
isbn = {9781450362290},
publisher = {Association for Computing Machinery},
address = {New York, NY, USA},
url = {https://doi.org/10.1145/3295500.3356159},
doi = {10.1145/3295500.3356159},
booktitle = {Proceedings of the International Conference for High Performance Computing, Networking, Storage and Analysis},
articleno = {76},
numpages = {19},
location = {Denver, Colorado},
series = {SC '19}
}

@inproceedings{lessonslearnedpmdbms,
author = {Wu, Yinjun and Park, Kwanghyun and Sen, Rathijit and Kroth, Brian and Do, Jaeyoung},
title = {Lessons Learned from the Early Performance Evaluation of Intel Optane DC Persistent Memory in DBMS},
year = {2020},
isbn = {9781450380249},
publisher = {Association for Computing Machinery},
address = {New York, NY, USA},
url = {https://doi.org/10.1145/3399666.3399898},
doi = {10.1145/3399666.3399898},
booktitle = {Proceedings of the 16th International Workshop on Data Management on New Hardware},
articleno = {14},
numpages = {3},
location = {Portland, Oregon},
series = {DaMoN '20}
}

@article{permabench,
author = {Benson, Lawrence and Papke, Leon and Rabl, Tilmann},
title = {PerMA-Bench: Benchmarking Persistent Memory Access},
year = {2022},
issue_date = {July 2022},
publisher = {VLDB Endowment},
volume = {15},
number = {11},
issn = {2150-8097},
url = {https://doi.org/10.14778/3551793.3551807},
doi = {10.14778/3551793.3551807},
journal = {Proc. VLDB Endow.},
month = {jul},
pages = {2463–2476},
numpages = {14}
}

@inproceedings{evaluationofinteloptane,
author = {Peng, Ivy B. and Gokhale, Maya B. and Green, Eric W.},
title = {System Evaluation of the Intel Optane Byte-Addressable NVM},
year = {2019},
isbn = {9781450372060},
publisher = {Association for Computing Machinery},
address = {New York, NY, USA},
url = {https://doi.org/10.1145/3357526.3357568},
doi = {10.1145/3357526.3357568},
booktitle = {Proceedings of the International Symposium on Memory Systems},
pages = {304–315},
numpages = {12},
location = {Washington, District of Columbia, USA},
series = {MEMSYS '19}
}

@article{evalindexes,
author = {Lersch, Lucas and Hao, Xiangpeng and Oukid, Ismail and Wang, Tianzheng and Willhalm, Thomas},
title = {Evaluating Persistent Memory Range Indexes},
year = {2019},
issue_date = {December 2019},
publisher = {VLDB Endowment},
volume = {13},
number = {4},
issn = {2150-8097},
url = {https://doi.org/10.14778/3372716.3372728},
doi = {10.14778/3372716.3372728},
journal = {Proc. VLDB Endow.},
month = {dec},
pages = {574–587},
numpages = {14}
}

@inproceedings{pmioprimitives,
author = {van Renen, Alexander and Vogel, Lukas and Leis, Viktor and Neumann, Thomas and Kemper, Alfons},
title = {Persistent Memory I/O Primitives},
year = {2019},
isbn = {9781450368018},
publisher = {Association for Computing Machinery},
address = {New York, NY, USA},
url = {https://doi.org/10.1145/3329785.3329930},
doi = {10.1145/3329785.3329930},
booktitle = {Proceedings of the 15th International Workshop on Data Management on New Hardware},
articleno = {12},
numpages = {7},
location = {Amsterdam, Netherlands},
series = {DaMoN'19}
}

@INPROCEEDINGS{gordon,
  author={Zhang, Jialiang and Beckwith, Nicholas and Li, Jing Jane},
  booktitle={2021 IEEE 29th Annual International Symposium on Field-Programmable Custom Computing Machines (FCCM)}, 
  title={GORDON: Benchmarking Optane DC Persistent Memory Modules on FPGAs}, 
  year={2021},
  volume={},
  number={},
  publisher={IEEE},
  pages={97-105},
  address={Orlando, FL, USA},
  doi={10.1109/FCCM51124.2021.00019}}

@ARTICLE{modelingnvm,
  author={Wang, Zixuan and Liu, Xiao and Yang, Jian and Michailidis, Theodore and Swanson, Steven and Zhao, Jishen},
  journal={IEEE Micro}, 
  title={Characterizing and Modeling Nonvolatile Memory Systems}, 
  year={2021},
  volume={41},
  number={3},
  pages={63-70},
  doi={10.1109/MM.2021.3065305}}

@inproceedings{pactree,
author = {Kim, Wook-Hee and Krishnan, R. Madhava and Fu, Xinwei and Kashyap, Sanidhya and Min, Changwoo},
title = {PACTree: A High Performance Persistent Range Index Using PAC Guidelines},
year = {2021},
isbn = {9781450387095},
publisher = {Association for Computing Machinery},
address = {New York, NY, USA},
url = {https://doi.org/10.1145/3477132.3483589},
doi = {10.1145/3477132.3483589},
booktitle = {Proceedings of the ACM SIGOPS 28th Symposium on Operating Systems Principles},
pages = {424–439},
numpages = {16},
location = {Virtual Event, Germany},
series = {SOSP '21}
}

@inproceedings {level_hashing,
author = {Pengfei Zuo and Yu Hua and Jie Wu},
title = {{Write-Optimized} and {High-Performance} Hashing Index Scheme for Persistent Memory},
booktitle = {13th USENIX Symposium on Operating Systems Design and Implementation (OSDI 18)},
year = {2018},
isbn = {978-1-939133-08-3},
address = {Carlsbad, CA},
pages = {461--476},
url = {https://www.usenix.org/conference/osdi18/presentation/zuo},
publisher = {USENIX Association},
month = oct,
}

@inproceedings {FASTFAIR,
author = {Deukyeon Hwang and Wook-Hee Kim and Youjip Won and Beomseok Nam},
title = {Endurable Transient Inconsistency in {Byte-Addressable} Persistent {B+-Tree}},
booktitle = {16th USENIX Conference on File and Storage Technologies (FAST 18)},
year = {2018},
isbn = {978-1-931971-42-3},
address = {Oakland, CA},
pages = {187--200},
url = {https://www.usenix.org/conference/fast18/presentation/hwang},
publisher = {USENIX Association},
month = feb,
}

@misc{redis,
	Title = {Redis.},
        author = {Redis Labs},
	url = {https://github.com/pmem/redis/tree/3.2-nvml},
	note = {Accessed: 2023-3-09},
        year = {2009}
 }

@misc{memcached,
	Title = {Memcached.},
        author = {Danga Interactive},
	url = {https://github.com/lenovo/memcached-pmem},
	note = {Accessed: 2023-3-09},
        year = {2003}
        }

@inproceedings{e9patch,
author = {Duck, Gregory J. and Gao, Xiang and Roychoudhury, Abhik},
title = {Binary Rewriting without Control Flow Recovery},
year = {2020},
isbn = {9781450376136},
publisher = {Association for Computing Machinery},
address = {New York, NY, USA},
url = {https://doi.org/10.1145/3385412.3385972},
doi = {10.1145/3385412.3385972},
booktitle = {Proceedings of the 41st ACM SIGPLAN Conference on Programming Language Design and Implementation},
pages = {151–163},
numpages = {13},
location = {London, UK},
series = {PLDI 2020}
}

@inproceedings{pmtest,
author = {Liu, Sihang and Wei, Yizhou and Zhao, Jishen and Kolli, Aasheesh and Khan, Samira},
title = {PMTest: A Fast and Flexible Testing Framework for Persistent Memory Programs},
year = {2019},
isbn = {9781450362405},
publisher = {Association for Computing Machinery},
address = {New York, NY, USA},
url = {https://doi.org/10.1145/3297858.3304015},
doi = {10.1145/3297858.3304015},
booktitle = {Proceedings of the Twenty-Fourth International Conference on Architectural Support for Programming Languages and Operating Systems},
pages = {411–425},
numpages = {15},
location = {Providence, RI, USA},
series = {ASPLOS '19}
}

@inproceedings{yat,
author = {Lantz, Philip and Dulloor, Subramanya and Kumar, Sanjay and Sankaran, Rajesh and Jackson, Jeff},
title = {Yat: A Validation Framework for Persistent Memory Software},
year = {2014},
isbn = {9781931971102},
publisher = {USENIX Association},
address = {USA},
booktitle = {Proceedings of the 2014 USENIX Conference on USENIX Annual Technical Conference},
pages = {433–438},
numpages = {6},
location = {Philadelphia, PA},
series = {USENIX ATC'14}
}

@inproceedings{jaaru,
author = {Gorjiara, Hamed and Xu, Guoqing Harry and Demsky, Brian},
title = {Jaaru: Efficiently Model Checking Persistent Memory Programs},
year = {2021},
isbn = {9781450383172},
publisher = {Association for Computing Machinery},
address = {New York, NY, USA},
url = {https://doi.org/10.1145/3445814.3446735},
doi = {10.1145/3445814.3446735},
booktitle = {Proceedings of the 26th ACM International Conference on Architectural Support for Programming Languages and Operating Systems},
pages = {415–428},
numpages = {14},
location = {Virtual, USA},
series = {ASPLOS '21}
}

@inproceedings {agamotto,
author = {Ian Neal and Ben Reeves and Ben Stoler and Andrew Quinn and Youngjin Kwon and Simon Peter and Baris Kasikci},
title = {{AGAMOTTO}: How Persistent is your Persistent Memory Application?},
booktitle = {14th USENIX Symposium on Operating Systems Design and Implementation (OSDI 20)},
year = {2020},
isbn = {978-1-939133-19-9},
pages = {1047--1064},
address = {Virtual},
url = {https://www.usenix.org/conference/osdi20/presentation/neal},
publisher = {USENIX Association},
month = nov,
}

@article{pmsematincs,
author = {Raad, Azalea and Wickerson, John and Neiger, Gil and Vafeiadis, Viktor},
title = {Persistency Semantics of the Intel-X86 Architecture},
year = {2019},
issue_date = {January 2020},
publisher = {Association for Computing Machinery},
address = {New York, NY, USA},
volume = {4},
number = {POPL},
url = {https://doi.org/10.1145/3371079},
doi = {10.1145/3371079},
journal = {Proc. ACM Program. Lang.},
month = {dec},
articleno = {11},
numpages = {31}
}

@inproceedings{woart,
author = {Se Kwon Lee and K. Hyun Lim and Hyunsub Song and Beomseok Nam and Sam H. Noh},
title = {{WORT}: Write Optimal Radix Tree for Persistent Memory Storage Systems},
booktitle = {15th USENIX Conference on File and Storage Technologies (FAST 17)},
year = {2017},
isbn = {978-1-931971-36-2},
address = {Santa Clara, CA},
pages = {257--270},
url = {https://www.usenix.org/conference/fast17/technical-sessions/presentation/lee-se-kwon},
publisher = {USENIX Association},
month = feb,
}

@inproceedings{xfdetector,
author = {Liu, Sihang and Seemakhupt, Korakit and Wei, Yizhou and Wenisch, Thomas and Kolli, Aasheesh and Khan, Samira},
title = {Cross-Failure Bug Detection in Persistent Memory Programs},
year = {2020},
isbn = {9781450371025},
publisher = {Association for Computing Machinery},
address = {New York, NY, USA},
url = {https://doi.org/10.1145/3373376.3378452},
doi = {10.1145/3373376.3378452},
booktitle = {Proceedings of the Twenty-Fifth International Conference on Architectural Support for Programming Languages and Operating Systems},
pages = {1187–1202},
numpages = {16},
location = {Lausanne, Switzerland},
series = {ASPLOS '20}
}

@misc{eadr,
	Title = {eADR: New Opportunities for Persistent Memory Applications},
        author = {Intel},
	url ={https://www.intel.com/content/www/us/en/developer/articles/technical/eadr-new-opportunities-for-persistent-memory-applications.html},
	note = {Accessed: 2023-03-28},
        year = {2021}
}

@misc{sql,
	Title = {Configure persistent memory (PMEM) for SQL Server},
	url ={https://learn.microsoft.com/en-us/sql/database-engine/configure-windows/configure-persistent-memory?view=sql-server-ver16},
        author = {Microsoft},
	note = {Accessed: 2023-03-28},
        year = {2023}
 
}

@inproceedings{ycsb,
 author = {Cooper, Brian F. and Silberstein, Adam and Tam, Erwin and Ramakrishnan, Raghu and Sears, Russell},
 title = {Benchmarking Cloud Serving Systems with YCSB},
 booktitle = {Proceedings of the 1st ACM Symposium on Cloud Computing},
 series = {SoCC '10},
 year = {2010},
 isbn = {978-1-4503-0036-0},
 location = {Indianapolis, Indiana, USA},
 pages = {143--154},
 numpages = {12},
 url = {http://doi.acm.org/10.1145/1807128.1807152},
 doi = {10.1145/1807128.1807152},
 acmid = {1807152},
 publisher = {ACM},
 address = {New York, NY, USA},
}

@inproceedings{pmdebugger,
    author = {Di, Bang and Liu, Jiawen and Chen, Hao and Li, Dong},
    booktitle = {Proceedings ofthe 26th ACM International Conference on
                 Architectural Support for Programming Languages and Operating
                 Systems},
    doi = {10.1145/3445814.3446744},
    isbn = {9781450383172},
    pages = {503--516},
    title = {{Fast, flexible, and comprehensive bug detection for persistent
             memory programs}},
    url = {https://doi.org/10.1145/3445814.3446744},
    year = {2021},
    series = {ASPLOS'21},
}

@inproceedings{witcher,
    author = {Fu, Xinwei and Kim, Wook-Hee and Shreepathi, Ajay Paddayuru and
              Ismail, Mohannad and Wadkar, Sunny and Lee, Dongyoon and Min,
              Changwoo},
    title = {Witcher: Systematic Crash Consistency Testing for Non-Volatile
             Memory Key-Value Stores},
    year = {2021},
    isbn = {9781450387095},
    publisher = {ACM},
    url = {https://doi.org/10.1145/3477132.3483556},
    doi = {10.1145/3477132.3483556},
    booktitle = {Proceedings of the ACM SIGOPS 28th Symposium on Operating
                 Systems Principles},
    pages = {100–115},
    numpages = {16},
    series = {SOSP'21},
}

@inproceedings{pmrace,
    author = {Chen, Zhangyu and Hua, Yu and Zhang, Yongle and Ding, Luochangqi},
    title = {Efficiently Detecting Concurrency Bugs in Persistent Memory
             Programs},
    year = {2022},
    isbn = {9781450392051},
    publisher = {ACM},
    url = {https://doi.org/10.1145/3503222.3507755},
    doi = {10.1145/3503222.3507755},
    booktitle = {Proceedings of the 27th ACM International Conference on
                 Architectural Support for Programming Languages and Operating
                 Systems},
    pages = {873–887},
    numpages = {15},
    series = {ASPLOS'22},
}

@inproceedings{durinn,
    author = {Xinwei Fu and Dongyoon Lee and Changwoo Min},
    title = {{DURINN}: Adversarial Memory and Thread Interleaving for Detecting
             Durable Linearizability Bugs},
    booktitle = {16th USENIX Symposium on Operating Systems Design and
                 Implementation},
    year = {2022},
    isbn = {978-1-939133-28-1},
    pages = {195--211},
    url = {https://www.usenix.org/conference/osdi22/presentation/fu},
    publisher = {USENIX Association},
    series = {OSDI'22},
}

@inproceedings{mumak,
    author = {Gon\c{c}alves, Jo\~{a}o and Matos, Miguel and Rodrigues, Rodrigo},
    title = {Mumak: Efficient and Black-Box Bug Detection for Persistent Memory},
    year = {2023},
    isbn = {9781450394871},
    publisher = {Association for Computing Machinery},
    address = {New York, NY, USA},
    url = {https://doi.org/10.1145/3552326.3587447},
    doi = {10.1145/3552326.3587447},
    booktitle = {Proceedings of the Eighteenth European Conference on Computer
                 Systems},
    pages = {734–750},
    numpages = {17},
    series = {EuroSys '23},
}

@inproceedings{hawkset,
    author = {Oliveira, Jo\~{a}o and Gon\c{c}alves, Jo\~{a}o and Matos, Miguel},
    title = {HawkSet: Automatic, Application-Agnostic, and Efficient Concurrent PM Bug Detection},
    year = {2025},
    isbn = {9798400711961},
    publisher = {Association for Computing Machinery},
    address = {New York, NY, USA},
    url = {https://doi.org/10.1145/3689031.3717477},
    doi = {10.1145/3689031.3717477},
    booktitle = {Proceedings of the Twentieth European Conference on Computer Systems},
    pages = {1092–1108},
    numpages = {17},
    location = {Rotterdam, Netherlands},
    series = {EuroSys '25}
}
\end{document}